\documentclass{PoS}

\usepackage[square,sort,comma,numbers]{natbib}

\def\aj{AJ}%
\def\araa{ARA\&A}%
\def\apj{ApJ}%
\def\apss{Ap\&SS}%
\def\aap{A\&A}%
\def\baas{BAAS}%
\def\mnras{MNRAS}%
\def\pasj{PASJ}%
\def\nat{Nature}%
\def\physscr{Phys.~Scr}%
\title{5 year Global 3-mm VLBI survey of Gamma-ray active blazars}

\ShortTitle{5 year Global 3-mm VLBI survey of Gamma-ray active blazars}

\author{\speaker{Jeffrey HODGSON}$^{1}$, Thomas P. Krichbaum$^{1}$, Alan P. Marscher$^{2}$, Svetlana G. Jorstad$^{2}$, Ivan Marti-Vidal$^{3}$, Michael Lindqvist$^{3}$, Michael Bremer$^{5}$, Salvador Sanchez$^{4}$, Pable de Vicente$^{6}$, Anton Zensus$^{1}$  \\
        (1) Max-Planck-Institut f\"{u}r Radioastronomie, Bonn, Germany\\
        (2) Boston University, Boston, United States \\
        (3) Onsala Space Observatory, Onsala, Sweden \\
        (4) Institut de Radioastronomie Millim\`{i}trique, Granada, Spain \\
        (5) Institut de Radioastronomie Millim\`{i}trique, Grenoble, France \\
        (6) Yebes Astronomical Observatory, Yebes, Spain \\
        E-mail: \email{jhodgson@mpifr-bonn.mpg.de}}
        
\abstract{The Global mm-VLBI Array (GMVA) is a network of 14 3\,mm and 7\,mm capable telescopes spanning Europe and the United States, with planned extensions to Asia. The array is capable of sensitive maps with angular resolution often exceeding 50\,$\mu$as. Using the GMVA, a large sample of prominent $\gamma$-ray blazars have been observed approximately 6 monthly from later 2008 until now. Combining 3\,mm maps from the GMVA with near-in-time 7\,mm maps from the VLBA-BU-BLAZAR program and 2\,cm maps from the MOJAVE program, we determine the sub-pc morphology and high frequency spectral structure of $\gamma$-ray blazars. The magnetic field strength can be estimated at different locations along the jet under the assumption of equipartition between magnetic field and relativistic particle energies. Making assumptions on the jet magnetic field configuration (e.g. poloidal or toroidal), we can estimate the separation of the mm-wave ``core'' and the jet base, and estimate the strength of the magnetic field there. The results of this analysis show that on average, the magnetic field strength decreases with a power-law $B \propto r^{-n}$, $n=0.3 \pm 0.2$.  This suggests that on average, the mm-wave ``core'' is $\sim 1-3$\,pc downstream of the de-projected jet apex and that the magnetic field strength is of the order $B_{\rm{apex}} \sim 5-20$\,kG, broadly consistent with the predictions of magnetic jet launching (e.g. via magnetically arrested disks (MAD)). }

\FullConference{12th European VLBI Network Symposium and Users Meeting\\
		 7-10 October 2014\\
		 Cagliari, Italy}

\begin{document}

\section{Introduction}

Strong magnetic fields have long been considered important in the production of AGN jets around super-massive black holes (SMBH) \citep{bz77}, and in the accretion disk \citep{balbus91}. Recently, there has been increasing observational evidence for particularly strong magnetic fields - near their theoretical maximum - under the model of magnetically arrested disks (MAD) \citep{koskesh,ghis14}. MAD describes a scenario where a very strong magnetic field is produced by accreting matter being dragged in a poloidal magnetic field, causing gas to fall onto the black hole with a higher-than-free-fall velocity \citep{bis74,nara03}. This generates a poloidal magnetic field which is then twisted around the black hole, creating a toroidal magnetic field in the jet \citep{tchek11,mckinney12}. \\

The Global mm-VLBI Array (GMVA) operates at 3\,mm and 7\,mm and can provide angular resolution exceeding 50\,$\mu$as at 86\,GHz \citep{hodgson14}. By combining GMVA observations at 3\,mm with near-in-time observations at 7\,mm as part of the VLBA-BU-BLAZAR program \cite{jor05}, we are able to spectrally decompose the VLBI structure of $\gamma$-ray bright blazars at different positions along the jet. This allows estimates of the magnetic field strength to be determined at these positions under the assumption of equipartition of magnetic field and relativistic particle energies. By making some further assumptions about the jet and magnetic field geometry, we can estimate the distance between the VLBI ``core'' (the most upstream visible part of the jet in VLBI maps) and the central SMBH. We can then extrapolate the magnetic field in the parsec-scale jet back to the SMBH. We find that the mm-wave ``core'' is located $\sim 1-3$\,pc downstream of the jet base and that the magnetic field strength is of the order $B_{\rm{apex}} \sim 5000-20000$\,G. Such magnetic field strengths are high but not inconsistent with those predicted by magnetic jet launching scenarios such as MAD. 

\section{Observations}

Six blazars are included in this analysis: 0716+714, 0836+710, 3C\,273, BL Lac, 3C\,454.3 and OJ\,287. Three millimetre VLBI data were observed during the period 2008.78-2012.35 using the Global mm-VLBI Array (GMVA) and CLEAN maps were produced using the Caltech DIFMAP package \citep{difmap}. Data were fringe-fitted and calibrated using standard procedures in AIPS for high frequency VLBI data reduction \citep[e.g.][]{jor05} with extended procedures written in ParselTongue as described by \citet{marti12}. Within AIPS, amplitudes were corrected for system temperatures, sky opacity, gain-elevation curves and then averaged over all IFs to increase SNR. Manual phase calibration was performed on the brightest sources and scans within the experiment. Data were correlated at the Max-Planck-Institut f\"{u}r Radioastronomie in Bonn, Germany. Complimenting GMVA observations are near-in-time observations from the use of 43 GHz VLBA data from the VLBA-BU-BLAZAR Monitoring Program  \href{http://www.bu.edu/blazars/VLBAproject.html}{VLBA-BU}, funded by NASA through the Fermi Guest Investigator Program. The VLBA is an instrument of the National Radio Astronomy Observatory. The National Radio Astronomy Observatory is a facility of the National Science Foundation operated by Associated Universities, Inc. The data were reduced using similar methods described by \citet{jor05}. 

\section{Analysis}

\subsection{``Core'' Identification}
\label{core_id}

A critical part of VLBI analysis is ``core'' identification, as it is required in order to perform kinematic analysis. The VLBI ``core'' is assumed to be the most upstream visible component in a VLBI image and is assumed to be stationary. The ``core'' is typically identified on the basis of i) morphology, ii) small size, higher flux densities and correspondingly higher brightness temperatures than downstream components iii) an optically thick (inverted) or flat spectrum and iv) increased levels of flux density variations. At wavelengths longer than at millimetre, the ``core'' is often relatively easy to identify, as it is usually the brightest component by far. However, at high frequencies, this can sometimes be difficult, as the sub-parsec scale structure can include quasi-stationary features that are sometimes brighter and more compact than the ``core'' \citep{jor05}. Nevertheless, based on the criteria listed above, it has been always possible to confidently identify the ``core'' in our VLBI maps. 

\subsection{Magnetic Fields}
\label{sec:mag_explain}

As VLBI at 3\,mm allows observations at or above the turnover frequency for synchrotron radiation, it allows for a novel approach for deriving estimates of the magnetic field in blazars. Although at least three spectral points are required to derive a spectrum (and determine with confidence the turnover frequency), the combined use of 3 and 7\,mm VLBI maps allows for limits to be computed for the magnetic field as a function of distance down the jet. In the future, the addition of 1\,mm and 2\,cm MOJAVE data will allow us to determine the turnover frequency more robustly. For this analysis, only components that could be fitted to a single, non-delta component at both frequencies were used. 

Jet emission is said to be in equipartition when relativistic particle and magnetic energies are equal. If we assume that the jet is in equipartition, we can compute an estimate of the minimum magnetic field strength can hence be calculated from (e.g. \cite{bach_05}):
\begin{equation}
B_{\rm{equi}} = 5.37 \times 10^{12} (S_{m}\nu_{m}D_{L}^{2}R^{-3})^{-2/7}\delta^{(2-2\alpha)/7} \rm{ [G]},
\end{equation}
where $D_{L}$ is the luminosity distance in Mpc, $\alpha$ is the spectral index, $\delta$ is the Doppler factor, $\nu_{m}$ is the turnover frequency for synchrotron emission in GHz, $S_{m}$ is the flux density at the turnover frequency in Jy and $R$ is the linear radius of the emitting region in cm at the turnover frequency. 
%\begin{equation}
%\kappa =  \frac{\Gamma \beta_{\rm{int}}\sin\theta_{0}(8\beta_{\rm{int}}^{2}\sin^{2}\theta_{0}-\Gamma^{-2})^{1/2}}{(1-\beta_{\rm{int}}^{2}\cos^{2}\theta_{0})^{1/2}},
%\end{equation} = (1-\beta_{\rm{int}}^{2}\cos^{2} \theta_{0})^{1/2} / [\Gamma \beta_{\rm{int}}\sin\theta_{0}(8\beta_{\rm{int}}^{2}\sin^{2}\theta_{0}-\Gamma^{-2})^{1/2}]$

\subsection{Distance to SMBH}
\label{sec:bh_dist}

The relativistic jet model was first proposed to explain the variability of non-thermal emission in quasars and blazars \citep{blandford_rees78,bk79}. As \citet{margt92} puts it:
\begin{quote}
The jet is assumed to be generated at some point $R_{\rm{apex}}$, beyond which it flows at a constant Lorentz factor $\Gamma$ (at speed $\beta c$), confined to a cone of constant opening half-angle $\phi$.
\end{quote}
If the jet is confined only by its own inertia, the jet density will decrease as $1/r^{2}$, where $r$ is the radial distance from the jet apex. If the magnetic field is parallel to the jet (toroidal), the magnetic field will decrease as $B_{\parallel} \propto r^{-1}$, when it is perpendicular (poloidal) $B_{\perp} \propto r^{-2}$. This can be re-expressed as $B \propto (1/r)^{n}$ \citep[e.g.][]{daly96}, where n is an exponent for toroidal (n=1) or poloidal (n=2) magnetic field configurations. We can compute an estimate of the distance from the ``core'' to the jet apex with the ratio of magnetic fields and the separation between the ``core'' (C1) and downstream component (C2):
\begin{equation}
\frac{B_{C1}}{B_{C2}} = \sqrt[n]{\frac{r_{\rm{C1}} + \Delta r_{C2}}{r_{\rm{C1}}}}
\end{equation}
Where $r_{\rm{C1}}$ is the distance to the black hole from the ``core'' in mas, $\Delta r_{C2}$ is the separation between the ``core'' and stationary feature. Rearranging, we get:
\begin{equation}
r_{\rm{C1}} = \frac{\Delta r_{C2}}{(B_{C1}/B_{C2})^{n}-1}.
\end{equation}
This can be generalised to be able to solve for the magnetic field at any arbitrary location along the jet:
\begin{equation}
\frac{B_{2}}{B_{1}} = \left[\frac{R_{1}}{R_{2}}\right]^{1/n},
\end{equation}
where $B_{1}$ and $B_{2}$ are the magnetic fields at arbitrary \emph{transverse} radii $R_{1}$ and $R_{2}$. Hence if the ``core'' is assumed to be resolved, we can estimate the magnetic field strength at a distance at any arbitrary distance (e.g. $10R_{S}$).
%Recent work by \citet{gabuzda14} and \citet{koskesh} suggests that toroidal (n=1) magnetic exist in AGN, although any other configuration could also exist. We can then estimate the magnetic field at the jet apex by extrapolating using the following equation:
%\begin{equation}
%B_{\Delta R} = B_{\rm{C1}} \left[ \frac{ r_{\rm{C1}}}{-\Delta r_{\rm{C1}-\Delta r} + \Delta r_{\rm{C1}} } \right]^{1/n} \rm{ [G]},
%\end{equation}
%where $B_{\rm{C1}}$ is the magnetic field in the ``core'' and $\Delta r_{\rm{C1}-\Delta r} $ is the difference between the ``core'' and the distance being solved for. Here, we solve for $\Delta r_{\rm{C1}-\Delta r} = 10R_{S}$. Where $R_{S}$ is the Schwarzschild radius.

\section{Discussion and Conclusions}

\begin{figure}[ht]
    \centering
    \includegraphics[width=0.8\linewidth]{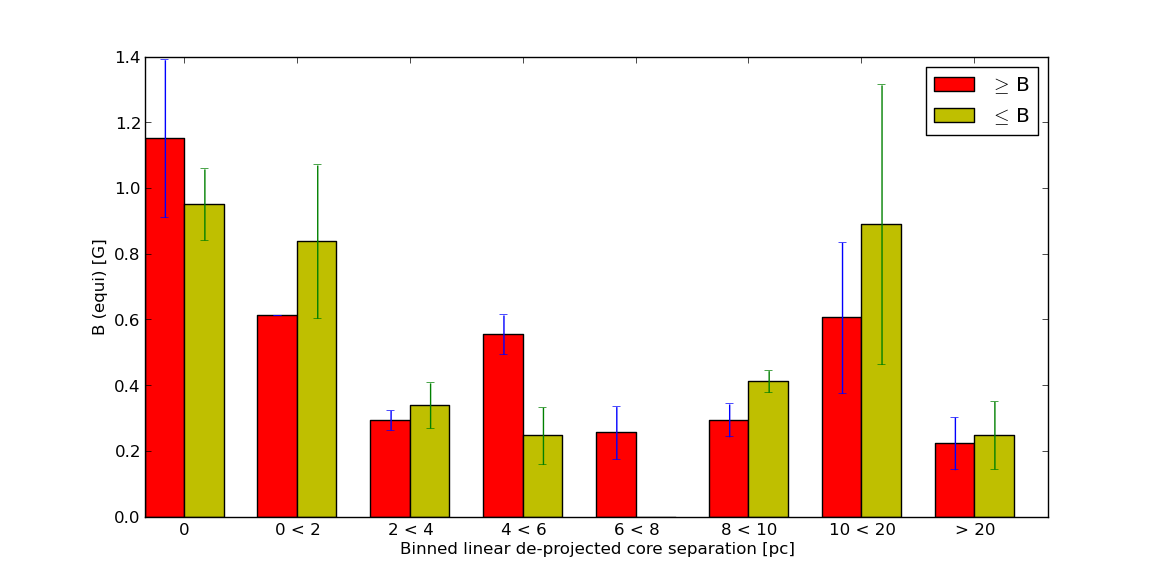}
    \caption[Binned B-fields (equi)]{Equipartition magnetic fields, binned by de-projected distance from the ``core''. Red bars indicate the average upper limits and yellow bars indicate the average lower limit. The error bars indicate the scatter of the limits. }
    \label{Bequi_binned}
\end{figure}

Although only upper and lower limits on the magnetic field strength can be computed, we can bin and average these limits. These averaged magnetic field limits are plotted in Fig. \ref{Bequi_binned} as a function of de-projected linear ``core'' separation and over all sources. De-projected distances are binned ever 2 parsecs over the first 10 pc. From this, we find that the Doppler corrected magnetic field strength in the ``core'' is on average between $\sim$0.8 and 1.4\,G, which is stronger than the results found via ``core-shift'' by \citet{os09}, which on a sample of 6 quasars found magnetic field strengths of 10s to 100s of mG. These values are not necessarily inconsistent with our results as these values were measured at cm-wavelengths and hence further downstream than the mm-wave ``core''. Our values are however consistent with the values found in the MOJAVE sample by \citet{push12}, which found values of $\sim$0.9\,G in FSRQs and $\sim$\,0.4\,G in BL Lacs. In our small sample of six sources, we find local maxima in the B-field distribution which may be interpreted as possibly due to recollimation and oblique shocks \citep[e.g.][]{laing_sheet}, but due to the low sample size, we cannot rule out any scenario. We can also test the rate at which the magnetic field decreases with increasing ``core'' separation, by performing a power-law fit to the limits, $B \propto r^{-n}$. This yields a value of $n=0.3 \pm 0.2$.  \\

Using the decrease in magnetic field strength of 0.9$< B <$1.4 at zero ``core'' separation to 0.2$< B <$0.7 at a de-projected separation of 4$-$6\,pc, yields an estimated distance from the jet apex to the mm-wave ``core'' of between $\sim$1-3\,pc. This likely places the mm-wave ``core'' outside of the broad line region. Taking a typical SMBH mass of $M_{\rm{BH}}=5\times10^{8}$ solar masses and assuming a toroidal (n=1) geometry, we can extrapolate the average jet magnetic field strengths back to a distance of 10\,$R_{S}$, yielding strengths of $B_{\rm{apex}} \sim 5-20$\,kG. This points towards a magnetic launching mechanism of the VLBI jets, as recently described in the MAD and other similar models \citep[e.g.][]{tchek11,mckinney12,mckinney14} and is near the maximum Eddington magnetic field \citep[e.g.][]{rees84}. An observational study by \citet{silan13} also found magnetic fields of the order thousands of Gauss in AGN, independently confirming such high values. \\

\citet{koskesh} showed that accretion disk luminosities are related to their black hole masses and hence to their magnetic field strengths. For black-hole masses of $> 10^{9}$ solar masses and higher accretion disk luminosities, the magnetic field estimates can be $B_{\rm{10R_{S}}} \sim$ 30\,kG or higher, consistent with the estimates derived here. However, if the magnetic field configuration is consistent with $n>1$, magnetic fields would be in the MG range, considerably stronger than expected from MAD and higher than theoretical Eddington limits, supporting the view that blazars have toroidal magnetic field configurations.

\section{Conclusions}

We have presented a novel approach to determining the magnetic field geometry of AGN jets. By spectrally decomposing 6 AGN using 3\,mm and 7\,mm VLBI, we have determined limits on the average magnetic field strength gradient. We have found that the average magnetic field strength in the mm-wave ``core'' is between $\sim$0.8 and 1.4\,G, decreasing to 0.3-0.4\,G at a de-projected ``core'' separation of 3-5\,pc. Fitting a power-law to these values ($B \propto r^{-n}$), yields $n=0.3 \pm 0.2$. By assuming a conically expanding jet and a toroidal (n=1) magnetic field geometry, we estimate the jet apex to be $\sim$1-3\,pc upstream of the mm-wave ``core'' and likely outside of the broad line region. Extrapolating the limits on the magnetic field strength to near the apex of the jet suggests magnetic field strengths of $B_{\rm{apex}} \sim 5-20$\,kG. We find that such strong magnetic fields are broadly consistent with magnetic jet launching, for example under the scenario of magnetically arrested discs.


\begin{thebibliography}{99}
\parskip0pt 
\itemsep0pt
  \bibitem[Bach et al.(2005)]{bach_05} Bach, U., Krichbaum, T.~P., Ros, E., et al.\ 2005, \aap, 433, 815 
  \bibitem[Balbus \& Hawley(1991)]{balbus91} Balbus, S.~A., \& Hawley, J.~F.\ 1991, \apj, 376, 214 ..
  \bibitem[Bisnovatyi-Kogan \& Ruzmaikin(1974)]{bis74} Bisnovatyi-Kogan, G.~S., \& Ruzmaikin, A.~A.\ 1974, \apss, 28, 45
  \bibitem[Blandford \& Znajek(1977)]{bz77} Blandford, R.~D., \& Znajek, R.~L.\ 1977, \mnras, 179, 433 .
  \bibitem[Blandford \& Rees(1978)]{blandford_rees78} Blandford, R.~D., \& Rees, M.~J.\ 1978, \physscr, 17, 265 
  \bibitem[Blandford \& K\"{o}nigl(1979)]{bk79} Blandford, R.~D., K\"{o}nigl, A.\ 1979, \apj, 232, 34 
  %\bibitem[Cawthorne(2006)]{Cawthorn06} Cawthorne, T.~V.\ 2006, \mnras, 367, 851 
  \bibitem[Cawthorne \& Cobb(1990)]{cc90} Cawthorne, T.~V., \& Cobb, W.~K.\ 1990, \apj, 350, 536 
  \bibitem[Gabuzda et al.(2014)]{gabuzda14} Gabuzda, D.~C., Cantwell, T.~M., \& Cawthorne, T.~V.\ 2014, \mnras, 438, L1 
  \bibitem[Ghisellini et al.(2014)]{ghis14} Ghisellini, G., Tavecchio, F., Maraschi, L., Celotti, A., \& Sbarrato, T.\ 2014, arXiv:1411.5368
  \bibitem[Guijosa \& Daly(1996)]{daly96} Guijosa, A., \& Daly, R.~A.\ 1996, \apj, 461, 600 
  \bibitem[Hodgson et al.(2014)]{hodgson14} Hodgson, J.~A., Krichbaum, T.~P., Marscher, A.~P., et al.\ 2014, arXiv:1407.8112 
  \bibitem[Jorstad et al.(2005)]{jor05} Jorstad, S.~G., Marscher, A.~P., Lister, M.~L., et al.\ 2005, \aj, 130, 1418 
  \bibitem[Laing(1980)]{laing_sheet} Laing, R.~A.\ 1980, \mnras, 193, 439 
  \bibitem[Marscher et al.(1992)]{margt92} Marscher, A.~P., Gear, W.~K., \& Travis, J.~P.\ 1992, Variability of Blazars, 85 
  \bibitem[Mart{\'{\i}}-Vidal et al.(2012)]{marti12} Mart{\'{\i}}-Vidal, I., Krichbaum, T.~P., Marscher, A., et al.\ 2012, \aap, 542, AA107 
  \bibitem[McKinney et al.(2012)]{mckinney12} McKinney, J.~C., Tchekhovskoy, A., \& Blandford, R.~D.\ 2012, \mnras, 423, 3083 
  \bibitem[McKinney et al.(2014)]{mckinney14} McKinney, J.~C., Tchekhovskoy, A., Sadowski, A., \& Narayan, R.\ 2014, \mnras, 441, 3177 
  \bibitem[Narayan et al.(2003)]{nara03} Narayan, R., Igumenshchev, I.~V., \& Abramowicz, M.~A.\ 2003, \pasj, 55, L69 
  \bibitem[O'Sullivan \& Gabuzda(2009)]{os09} O'Sullivan, S.~P., \& Gabuzda, D.~C.\ 2009, \mnras, 400, 26 
  \bibitem[Pushkarev et al.(2012)]{push12} Pushkarev, A.~B., Hovatta, T., Kovalev, Y.~Y., et al.\ 2012, \aap, 545, AA113 
  \bibitem[Rees(1984)]{rees84} Rees, M.~J.\ 1984, \araa, 22, 471 
  \bibitem[Shepherd et al.(1994)]{difmap} Shepherd, M.~C., Pearson, T.~J., \& Taylor, G.~B.\ 1994, \baas, 26, 987 
  \bibitem[Silant'ev et al.(2013)]{silan13} Silant'ev, N.~A., Gnedin, Y.~N., Buliga, S.~D., Piotrovich, M.~Y., \& Natsvlishvili, T.~M.\ 2013, Astrophysical Bulletin, 68, 14 
  \bibitem[Tchekhovskoy et al.(2011)]{tchek11} Tchekhovskoy, A., Narayan, R., \& McKinney, J.~C.\ 2011, \mnras, 418, L79 
  \bibitem[Zamaninasab et al.(2014)]{koskesh} Zamaninasab, M., Clausen-Brown, E., Savolainen, T., \& Tchekhovskoy, A.\ 2014, \nat, 510, 126 
\end{thebibliography}
\end{document}